\documentclass[final,5p]{elsarticle}

\usepackage{graphicx}
\usepackage{color}
\usepackage{amssymb}
\usepackage{amsmath}
\usepackage{braket}
\usepackage{hyperref}
\usepackage{bm}

\hypersetup{bookmarksnumbered,%
	colorlinks,%
	linkcolor=blue,%
	citecolor=blue,%
	plainpages=false,%
	pdfstartview=FitH}

\newcommand{\nc}{\newcommand}           
\nc{\vc}[1]     {\mbox{\boldmath $#1$}} 
\nc{\mapleft}[1]{                       
 \smash{\mathop{                        %
  \hbox to 0.90cm{\rightarrowfill} }\limits_{#1}}}
\nc{\figwidth}{0.45\textwidth}          

\journal{Physics Letters B}

\begin{document}

\begin{frontmatter}
\title{High-momentum components in the $^{4}$He nucleus caused by inter-nucleon correlations}

\author[1]{Mengjiao Lyu}\ead{mengjiao@rcnp.osaka-u.ac.jp}

\author[2,1]{Takayuki Myo\corref{cor1}}
\cortext[cor1]{Corresponding author}\ead{takayuki.myo@oit.ac.jp}

\author[1]{Hiroshi Toki}

\author[1]{Hisashi Horiuchi}

\author[3]{Chang Xu}

\author[3]{Niu Wan} 

\address[1]{Research Center for Nuclear Physics (RCNP), Osaka
University, Ibaraki, Osaka 567-0047, Japan}
\address[2]{General Education, Faculty of Engineering, Osaka
Institute of Technology, Osaka, Osaka 535-8585, Japan}
\address[3]{School of Physics, Nanjing University,
  Nanjing 210093, China}



\begin{abstract}
High-momentum components of nuclei are essential for understanding the
underlying inter-nucleon correlations in nuclei. We perform the comprehensive
analysis for the origin of the high-momentum components of $^{4}$He in the
framework of Tensor-optimized High-momentum Antisymmetrized Molecular Dynamics
(TO-HMAMD), which is a completely variational approach as an {\it ab initio}
theory starting from the bare nucleon-nucleon interaction. The analytical
derivations are provided for the nucleon momentum distribution of the
Antisymmetrized Momentum Dynamics (AMD) wave functions, with subtraction of
center-of-mass motion. The nucleon momentum distribution for $^{4}$He is
calculated by applying a new expansion technique to our {\it ab initio} wave
function, and agrees with the values extracted from experimental data up to
the high-momentum region. Fine-grained analysis is performed for the
high-momentum components in $^{4}$He with respect to different nucleon
correlations. Contributions from tensor, central with short-range, and
many-body correlations are extracted from the nucleon momentum distributions.
The manifestation of tensor correlation around 2 fm$^{-1}$ region is
explicitly confirmed by comparing the momentum distributions predicted using
different types of $NN$ interactions with and without the tensor force.
\end{abstract}




\end{frontmatter}

\section{Introduction} 
In atomic nuclei, the high-momentum components provide an important window for
understanding the underlying inter-nucleon correlations induced by the bare
nucleon-nucleon ($NN$) interactions \cite{pieper01b}. In some studies, the
high-momentum components are predicted to contribute about 20\% to the total
wave function in both finite nuclei \cite{hen14, ciofi15, hen17} and nuclear
matter \cite{xu13, oertel17}. In the reproduction of the high-momentum
components, it is essential to describe correctly the strong $NN$-correlations
in the nuclear wave function, which is the major difficulty of the {\it ab
initio} calculations for nuclear systems. For the $NN$ correlations, it is
well known that both tensor and short-range correlations are induced by the
attraction from tensor force and the short-range repulsion respectively, in
the bare interaction \cite{pieper01}. Accordingly, many-body wave functions
are built upon the $NN$-correlation functions. In the Green's Function Monte
Carlo (GFMC) method \cite{pieper01}, a Jastrow type of correlation function is
introduced in the trial wave function, and is assumed to be a product form of
many-kinds of $NN$-correlation functions. In Ref.~\cite{neff03}, the unitary
operators are introduced for the inter-nucleon correlations of tensor and
short-range types.

Recently, the ``Tensor-optimized Antisymmetrized Molecular Dynamics'' (TOAMD)
method is developed in Refs.~\cite{myo15, myo17a, myo17b, myo17c, myo17d} by
multiplying the $NN$-correlations into the ``Antisymmetrized Molecular
Dynamics'' (AMD) reference wave function, which itself is a microscopic
framework successfully applied in the study of nuclear cluster states
\cite{kanada03,kanada12}. Both the tensor correlation function $F_{D}$ and the
central correlation function $F_{S}$ are introduced in the TOAMD wave
function. By including up to the second orders in the cluster expansion,
$i.e.$ $F_{D,S}$ and $F_{D,S}^{2}$, the TOAMD wave function reproduces well
the binding energies and radii of $s$-shell nuclei using the AV8$^\prime$ bare
interaction \cite{myo17a}. In later works, it is also found that the tensor
\cite{myo17e} and short-range \cite{myo18} correlations are well described
using the high-momentum $NN$ pairing technique \cite{itagaki18}, which is
named as the ``High-momentum Antisymmetrized Molecular Dynamics'' (HM-AMD) in
Refs.~\cite{myo17e, myo18}. This technique has been applied to the
calculations of both finite nuclei \cite{myo17e, myo18, itagaki18, matsuno18,
zhao19} and nuclear matters \cite{myo19, wan19}. Hybridizing the TOAMD and
HM-AMD methods, the ``Tensor-optimized High-momentum Antisymmetrized Molecular
Dynamics'' (TO-HMAMD) approach is formulated, which also provides the same
quality of the nuclear properties of $^{3}$H \cite{lyu18} and $^{4}$He
\cite{lyu18b} as other {\it ab initio} calculations in TOAMD or GFMC
frameworks. In the methods of TOAMD, HM-AMD, and TO-HMAMD, the wave functions
are variationally determined by minimizing the total energies, and the
high-momentum components are produced naturally in the optimized wave
function. In the TOAMD wave function, the $NN$ correlations are described by
using the variational correlation functions. The tensor correlation $F_{D}$ is
inherited from the preceding ``Tensor-optimized Shell Models'' (TOSM) approach
which provides good descriptions for the light nuclei \cite{myo05, myo07,
myo07_11, myo09, myo11}. In Ref.~\cite{myo11}, it is found that the spatial
shrinkage of particle states is essential for the tensor contribution, which
is further discussed in Ref.~\cite{itagaki18} and related to the emergence of
high-momentum components in nuclei. 

Experimentally, the high-momentum components are probed by using the electron
scattering \cite{hen14, ciofi15, hen17, subedi08, jefferson, clas18,
cruztorres19, pybus20} and the proton induced reactions \cite{tang03,
piasetzky06, terashima18}. In Ref.~\cite{subedi08}, the tensor correlation is
suggested by the prominent population of the proton-neutron pairs in the
high-momentum region. More recent developments are introduced in review papers
\cite{ciofi15, hen17}. 

Theoretically, the momentum distributions of light nuclei have been calculated
using bare interactions \cite{schiavilla86, morita88, schiavilla07, wiringa08,
alvioli12, alvioli13, wiringa14, neff15, ryckebusch19}. For $^{4}$He nucleus,
the origin of the high-momentum components has been discussed in various
studies. In Ref.~\cite{morita88}, the momentum distribution is calculated
using Reid $NN$ force and the contributions from $S$- and $D$-wave components
are discussed, while the binding energy of $^{4}$He is not reproduced in this
reference. In Ref.~\cite{wiringa14}, the proton momentum distributions are
compared for the central and central+tensor channels of the Variational Monte
Carlo wave function. In Refs.~\cite{alvioli13, neff15}, the spin-isospin
decomposition is performed for the proton momentum distribution. The
decomposition regarding different kinds of $NN$ pairs is provided in
Refs.~\cite{alvioli12, ryckebusch19}. However, the inter-nucleon correlations
become complex in many-body system and then the physical origins of
high-momentum components has not been clarified in relation to the
inter-nucleon correlations, such as the tensor or short-range ones. For
example, the nucleon momentum distribution excited by the short-range
repulsion can be coupled to the uncorrelated term in the reference $0s$-state
in $^{4}$He. In addition, the contributions from many-body correlations beyond
the two-body case, which correspond to the second or higher orders of cluster
expansion, is unknown. Hence, it is worthwhile to perform the thorough
investigations on the origins of the high-momentum components with respect to
different types of inter-nucleon correlations. 

In the present work, we provide a general method to calculate the momentum
distribution of nuclei with our {\it ab initio} approach of TO-HMAMD
\cite{lyu18b}. Using this method, we perform the fine-grained analysis for the
physical origins of high-momentum components in the $^{4}$He nucleus by the
decomposition of the {\it ab initio} wave function into the AMD, central,
tensor and many-body channels, which provides a different aspect of the
inter-nucleon correlations as compared to pioneer works. We also investigate
the importance of the tensor correlation in the high-momentum region by
comparing the results using different $NN$ interactions with and without the
tensor force.

\section{Formulation}
We describe the nucleon momentum distribution of the $^{4}$He nucleus in the
framework of TO-HMAMD. In Sec.~\ref{subsec:wf} we introduce the formulation of
TO-HMAMD wave function. In Sec.~\ref{subsec:amd}, we provide the first
formulation for the nucleon momentum distribution of the (HM-)AMD wave
functions. In Sec.~\ref{subsec:projection}, we introduce a projection approach
to expand the {\it ab initio} TO-HMAMD wave function by using the HM-AMD
bases, for the calculation of momentum distributions. 

\subsection{The {\it ab initio} wave function of $^{4}$He}
\label{subsec:wf}
We recapitulate the theoretical framework for the {\it ab initio} wave
function used in this work. Detailed explanations are found in
Ref.~\cite{lyu18b} and references there in.

In the coordinate space, the (HM-)AMD wave function with mass number $A$ is
written as
\begin{equation}
\ket{\Psi_{\text{(HM)AMD}}}=
  \mathcal{A}\left\{
    \prod_{i=1}^{A} \psi_{\alpha_{i}}(\bm{r}_{i},\bm{Z}_{i})
  \right\},
\end{equation}
where $\mathcal{A}$ is the antisymmetrizer and the nucleon state
$\psi_{\alpha}$ is given as the Gaussian wave packet 
\begin{equation}\label{eq:wavepacket}
  \psi_{\alpha}(\bm{r},\bm{Z})\propto
  e^{-\nu(\bm{r}-\bm{Z})^{2}} \chi_{\alpha}.
\end{equation}
Here $\alpha$ denotes the spin and isospin of the nucleon state and the range
parameter $\nu$ is determined in the energy variation. Coordinate $\bm{Z}$ is
the centroid of Gaussian wave packets which is typically real satisfying the
condition of $\sum_{i}^{A}\bm{Z}_{i}=\bm{0}$. For HM-AMD wave functions, the
imaginary shifts $\pm i\bm{D}$ are additionally introduced in a pairwise form
for two nucleons among $A$-nucleons as
\begin{equation}\label{eq:shift}
  \begin{split}
      \bm{Z}_i &\rightarrow \bm{Z}_i + i \bm{D},\\
      \bm{Z}_j &\rightarrow \bm{Z}_j - i \bm{D},
  \end{split}
\end{equation}
where the imaginary shift $\bm{D}$ represents the momentum component of the
Gaussian wave packet in Eq.~(\ref{eq:wavepacket}) such as
$\braket{\bm{k}}=2\nu\bm{D}$. From this property, using a large magnitude of
$\bm{D}$, we can describe the high-momentum components induced by the $NN$
correlations. While Eq.~(\ref{eq:shift}) describes only single $NN$ pair using
$i\bm{D}$, another pair with different $i\bm{D}'$ can be successively added in
a HM-AMD basis to describe simultaneously the correlations induced by two $NN$
pairs \cite{myo18,lyu18b}. The wave functions that include HM-AMD bases of up
to single or double $NN$ pairs are named as ``Single HMAMD'' and ``Double
HMAMD'', respectively. The TO-HMAMD wave function is then formulated using
correlation functions $F_{D}$ and $F_{S}$ as
\begin{equation}
|\Psi_{\rm TO-HMAMD} \rangle 
  =  (1+F_{D}+F_{S}) \ket{\Psi_{{\rm HM-AMD}}},    \label{eq:to-hmamd}
\end{equation}
where 
\begin{align}
&{F}_{D}= \sum_{i<j}^A\sum_{t} f_{D}^{t}(r_{ij}) 
    O_{ij}^t  r_{ij}^{2} S_{12}(\hat{r}_{ij}),\label{eq:fd}\\
&{F}_{S}= \sum_{i<j}^A\sum_{t,s} f_{S}^{t,s}(r_{ij})
    O_{ij}^t O_{ij}^s. \label{eq:fs}
\end{align}
Here, ${F}_{D}$ are the tensor correlation operators and ${F}_{S}$ are the
central correlation operators, with the operators $O_{ij}^t=(\bm \tau_i \cdot
\bm \tau_j)^t$ and $O_{ij}^s=(\bm \sigma_i \cdot \bm \sigma_j)^s$.
Superscripts $s$ and $t$ are used to distinguish the spin and isospin
channels. The pair functions $f^t_D$ and $f^{t,s}_S$ are determined in the
minimization of the total energy. When imaginary shifts $i\bm{D}$ are not
included in Eq.~(\ref{eq:to-hmamd}), the wave function reduces to the TOAMD
wave function of the first order \cite{myo15}
\begin{equation}
|\Psi_{\rm TOAMD} \rangle 
  =  (1+F_{D}+F_{S}) \ket{\Psi_{{\rm AMD}}}.    \label{eq:toamd}
\end{equation}
Using the TO-HMAMD wave function in Eq.~(\ref{eq:to-hmamd}), we obtain the
total energy as $-$24.74 MeV and the radius as 1.51 fm for the $^{4}$He
nucleus with AV8$^\prime$ bare interaction \cite{lyu18b}, which reproduce the
GFMC results \cite{pieper01}. From these results, we expect that the
high-momentum components are accurately described by the TO-HMAMD wave
function as an {\it ab initio} approach.

\subsection{Momentum distribution for (HM-)AMD wave function}
\label{subsec:amd}
The (HM-)AMD wave function in the momentum space $\Phi$ can be written as
\begin{equation}\label{eq:amd-k}
\Phi^{\text{(HM-)AMD}}\left(\bm{k}_{1}, \cdots, \bm{k}_{A}\right)=
  \mathcal{A}\left\{
    \prod_{i=1}^{A} \phi_{\alpha_{i}}(\bm{k}_{i},\bm{Z}_{i})
  \right\},
\end{equation}
where $\bm{k}_{i}$ denotes a momentum of each nucleon in the laboratory frame,
and $\phi_{\alpha}(\bm{k},\bm{Z})$ is the Fourier transformation of the
nucleon wave function $\psi_{\alpha}(\bm{r},\bm{Z})$ in
Eq.~(\ref{eq:wavepacket}). The (HM-)AMD wave function has a center-of-mass
component $\Phi_{\textrm{G}}(\bm{k}_{\textrm{G}})$ which can be factorized in
Eq.~(\ref{eq:amd-k}) as
\begin{equation}\label{eq:com}
\Phi =
  \Phi_{\textrm{I}} \cdot \Phi_{\textrm{G}}\left(\bm{k}_{\textrm{G}}\right),
\end{equation}
where $\Phi_{\textrm{I}}$ is the internal wave function and the center-of-mass
momentum is given as $\bm{k}_{\textrm{G}}=\sum_{i=1}^{A}\bm{k}_{i}$ and
\begin{equation}
\Phi_{\textrm{G}}\left(\bm{k}_{\textrm{G}}\right) =
  \frac{1}{(2 \pi A \nu)^{3 / 4}} e^{-\bm{k}_{\textrm{G}}^{2} /(4 A \nu)}.
\end{equation}
The operator of the nucleon momentum distribution is defined as
\begin{equation}
\hat{n}(\bm{k})\equiv
  \sum_{i=1}^{A}\delta\left(\bm{b}_{i}-\bm{k}\right),
\end{equation}  
where $\bm{b}_{i}$ is a nucleon momentum in the center-of-mass frame
satisfying $\bm{k}_{i}=\bm{b}_{i}+\bm{k}_{\textrm{G}}/A$ and $\sum_{i=1}^{A}
\bm{b}_{i}=\bm{0}$. Similarly, we define the nucleon momentum distribution
operator in the laboratory frame as
\begin{equation}\label{eq:rhog}
\hat{n}_{\textrm{G}}(\bm{k})\equiv
  \sum_{i=1}^{A}
  \delta\left(\bm{k}_{i}-\bm{k}\right).
\end{equation}
The expectation value of $\hat{n}_{\textrm{G}}(\bm{k})$ can be calculated
directly using the (HM-)AMD wave function as
\begin{equation}
n_{\textrm{G}}(\bm{k})
  =\frac{\braket{\Phi|\widehat{n}_{\textrm{G}}| \Phi}}
      {\braket{\Phi | \Phi}}
  =\frac{1}{(2 \pi)^{3}} 
    \int d \bm{r} e^{-i \mathbf{k} \cdot \mathbf{r}}       
    \cdot \widetilde{n}_{\textrm{G}}(\bm{r}),
\end{equation}
where
\begin{equation}\label{eq:ngr}
\widetilde{n}_{\textrm{G}}(\bm{r})
  =\sum_{i=1}^{A} 
    \frac{\braket{\Phi|e^{i \bm{k}_{i} \cdot \bm{r}}| \Phi}}
    {\braket{\Phi | \Phi}}
=\sum_{i, j=1}^{A}\braket{\phi_{i}|e^{i \bm{k} \cdot \bm{r}}| \phi_{j}}    
    \cdot B_{j i}^{-1}.
\end{equation}
We note that in Eq.~(\ref{eq:ngr}) the integrations are performed in momentum
space over momentum $\bm{k}$, and $B_{ij}=\braket{\phi_{i}|\phi_{j}}$ is the
overlap matrix of single nucleon states where we simply write it without
$\alpha$. On the other hand, the expectation value of $\hat{n}_{\rm
G}(\bm{k})$ can be given using $\bm{b}_{i}$ and the relation in
Eq.~(\ref{eq:com}) as
\begin{equation}\label{eq:rhog2}
\begin{aligned} 
\widetilde{n}_{\rm G}(\boldsymbol{r}) 
& \equiv \sum_{i=1}^{A} 
  \frac{
    \braket{
      \Phi|
        e^{i\left(\boldsymbol{b}_{i}+\boldsymbol{k}_{G} / A\right) 
        \cdot \boldsymbol{r}}
      |\Phi}}
  {
    \langle\Phi | \Phi\rangle}\\
&=\sum_{i=1}^{A} 
  \frac{
    \braket{\Phi_{\textrm{I}}|
      e^{i \boldsymbol{b}_{i} \cdot \boldsymbol{r}}
    |\Phi_{\textrm{I}}}}
  {
    \braket{\Phi_{\textrm{I}} | \Phi_{\textrm{I}}}} 
\cdot 
  \frac{
    \braket{\Phi_{\textrm{G}}|
      e^{i \boldsymbol{k}_{G} \cdot \boldsymbol{r} / A}
    |\Phi_{\textrm{G}}}}
  {
    \braket{\Phi_{\textrm{G}} | \Phi_{\textrm{G}}}},
\end{aligned}
\end{equation}
where the center-of-mass term is given as
\begin{equation}
  \frac{
      \braket{\Phi_{\textrm{G}}|
        e^{i \boldsymbol{k}_{G} \cdot \boldsymbol{r} / A}
      |\Phi_{\textrm{G}}}}
    {
      \braket{\Phi_{\textrm{G}} | \Phi_{\textrm{G}}}}
  =
    \exp \left[-\frac{\nu}{2 A} \boldsymbol{r}^{2}\right].
\end{equation}
The nucleon momentum distribution in the center-of-mass frame can now be
expressed using Eq.~(\ref{eq:rhog2}) as
\begin{equation}\label{eq:rho}
\begin{aligned}
n(\bm{k})
=&\frac{\braket{\Phi_{\textrm{I}}|\widehat{n}| \Phi_{\textrm{I}}}}
  {\braket{\Phi_{\textrm{I}} | \Phi_{\textrm{I}}}}\\
=&\frac{1}{(2\pi)^{3}}\int d \bm{r} e^{-i\bm{k}\cdot\bm{r}}
  \cdot \widetilde{n}_{G}(\bm{r})
  \cdot\exp\left( 
    \frac{\nu\bm{r}^{2}}{2A}  
  \right).
\end{aligned}
\end{equation}
Substituting Eq.~(\ref{eq:ngr}) into Eq.~(\ref{eq:rho}) we have the momentum
distribution in (HM-)AMD
\begin{equation}\label{eq:amd-momentum}
\begin{aligned}
&n(\bm{k})
=
  \left(\frac{1}{2\pi\nu\epsilon}\right)^{3/2}\\
&\qquad\quad\times 
  \sum_{i,j=1}^{A}\exp\left[
      -\frac{1}{2\nu\epsilon}\left(
          \bm{k}-i\nu(\bm{Z}^*_{i}-\bm{Z}_{j})
        \right)^{2}
  \right]B_{ij}B^{-1}_{ji},
\end{aligned}
\end{equation}
where $\epsilon=(A-1)/A$. We note that for HM-AMD wave functions, the centroid
parameters $\bm{Z}_{i,j}$ in Eq.~(\ref{eq:amd-momentum}) are complex with
imaginary shifts. The momentum distribution satisfies the normalization
condition 
\begin{equation}
  \int d\bm{k}\, n(\bm{k})=A.
\end{equation}
We note that the equations above are formulated in general for nucleus with
any mass number $A$. In addition, the center-of-mass motion is completely
subtracted  in this formulation, which is an significant advantage of this
research. When the total wave function
$\ket{\Phi}=\sum_{a}c_{a}\ket{\Phi_{a}}$ is a superposition of the AMD bases
$\ket{\Phi_{a}}$, the corresponding nucleon momentum distribution is given as
\begin{equation}\label{eq:rho-expan}
  \begin{aligned}
    n(\bm{k})
=&
    \frac{1}
      {\braket{\Phi|\Phi}}
    \sum_{a,b}
      c_{a}^{*}c_{b}\braket{\Phi_{a}|\widehat{n}|\Phi_{b}}.\\
  \end{aligned}
\end{equation}
In (HM-)AMD, usually the angular momentum projection \cite{horiuchi86} is
adopted to restore the rotational symmetry, which is mathematically a
superposition of rotated (HM-)AMD bases, and hence the corresponding nucleon
momentum distribution of $\ket{\Phi^{JM}}$ can be calculated using
Eq.~(\ref{eq:rho-expan}) for the state with total spin $J$ and $z$-component
$M$.

\subsection{Expansion of TO-HMAMD wave function}
\label{subsec:projection}
To calculate the nucleon momentum distribution of the {\it ab initio} TO-HMAMD
wave function $\ket{\Psi}$, we expand the $\ket{\Psi}$ in coordinate space by
a set of HM-AMD bases $\{\ket{\Psi_{1}},\ket{\Psi_{2}},\dots,\ket{\Psi_{n}}\}$
with the number of $n$ as
\begin{equation}\label{eq:expan-orgional}
    \ket{\Psi}\approx
    \sum_{i=1}^{n}C_{i}\ket{\Psi_{i}},
\end{equation}
where $C_{i}$ are expansion coefficients. The momentum distribution can be
calculated using HM-AMD basis with Eq.~(\ref{eq:rho-expan}). To obtain the
coefficients $C_{i}$, we first construct a new set of the orthonormal bases
$\ket{\widetilde{\Psi}_{k}}$ in the linear combination of the non-orthogonal
HM-AMD basis states, as
\begin{equation}\label{eq:orthbases}
\ket{\widetilde{\Psi}_{k}}
  =\sum_{j}\widetilde{v}_{j,k}\ket{\Psi_{j}}.
\end{equation}
The {\it ab initio} wave function $\ket{\Psi}$ can be expanded with
$\ket{\widetilde{\Psi}_{k}}$ as
\begin{equation}\label{eq:orthexpan}
\ket{\Psi}\approx\sum_{k} \mathcal{P}_{k} \ket{\widetilde{\Psi}_{k}},
\end{equation}
where
\begin{equation}
\mathcal{P}_{k}=
  \braket{\widetilde{\Psi}_{k}|\Psi}.
\end{equation}
Using Eqs.~(\ref{eq:orthbases}) and (\ref{eq:orthexpan}), the coefficients
$C_{i}$ are obtained as
\begin{equation}
    C_{i}=\sum_{k} \widetilde{v}_{i,k}\mathcal{P}_{k}
\end{equation}

In realistic calculations, the number $n$ of the HM-AMD bases used in
Eq.~(\ref{eq:expan-orgional}) is finite and the quality of this expansion can
be estimated using the overlap 
\begin{equation}
\mathcal{O}=
    \braket{\Psi| \sum_{k} \mathcal{P}_{k}{\widetilde{\Psi}_{k}}}=  
  \sum_{k}\mathcal{P}_{k}^{2}.
\end{equation}
For accurate expansion, the overlap $\mathcal{O}$ should be close to unity. 

\section{Results}
\begin{figure}[htb]
    \centering
    \includegraphics[width=\figwidth]{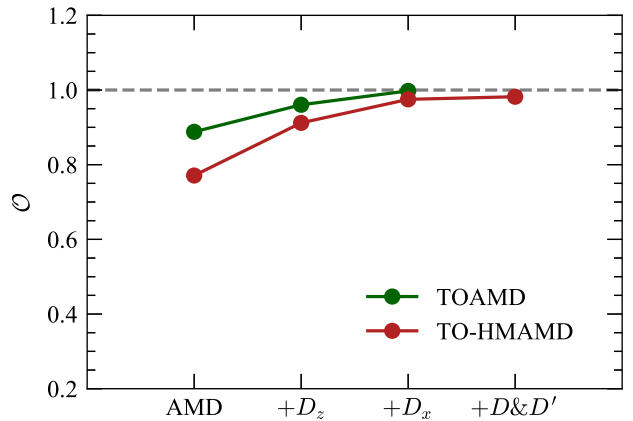}
    \caption{Overlap $\mathcal{O}$ between the {\it ab initio} wave function
    and its expansion with respect to the successive additions of HM-AMD
    bases. Notations are the same as used in Table \ref{tab:overlap}.}
    \label{fig:overlap}
\end{figure}

We first check the overlap $\mathcal{O}$ by successively enlarging the
functional space expanded by the HM-AMD bases, as shown in Table
\ref{tab:overlap} and Fig.~\ref{fig:overlap}. The setups of the Single HMAMD
bases are the same as in Ref.~\cite{lyu18b}, and the Double HMAMD bases are
additionally included. The overlap value $\mathcal{O}=99.8\%$ is obtained for
the TOAMD wave function, which confirms the accuracy and reliability of this
expansion method \cite{lyu18b}. It is found that when the Single HMAMD bases
are included, the overlap $\mathcal{O}$ for the TO-HMAMD wave function is
$\mathcal{O}=97.5\%$, almost unity. This is surprising because it is known
that the second order of the $NN$ correlations is important to converge the
solutions of the {\it ab initio} wave functions of nuclei \cite{myo17a} and
they are expected to be described using Double HMAMD bases \cite{lyu18b}. To
understand this behavior, we note that the Single HMAMD bases are not
orthogonal to the Double HMAMD bases, hence the effects of the second order
terms are partially included in the expansion within the Single HMAMD
functional space. In this work, we perform the expansion within Single HMAMD
bases so as to get the sufficient numerical accuracy for the momentum
distribution of nuclei.

\begin{table}[htb]
    \begin{center}
    \caption{Overlap $\mathcal{O}$ between the {\it ab initio} wave function
    and its expansion with respect to the successive additions of HM-AMD
    bases. ``$+D_{z}$'' and ``$+D_{x}$'' denote the additions of Single HM-AMD
    bases with imaginary shifts in $z$- and $x$-directions, respectively.
    ``$+D\&D'$'' denotes the addition of Double HMAMD bases. The total number
    of HM-AMD bases used for expansion is $n$.}
    \label{tab:overlap}
    \begin{tabular}{p{24mm}p{10mm}p{10mm}p{10mm}p{12mm}}
    \noalign{\hrule height 0.5pt}
              &AMD        & $+D_{z}$  & $+D_{x}$  &$+D\&D'$\\
    \noalign{\hrule height 0.5pt}
$n$           & 1         & 49        & 97        & $\sim$1000 \\
$\mathcal{O}(\textrm{TOAMD})$ & 0.888     & 0.960     & 0.998     &    \\
$\mathcal{O}(\textrm{TO-HMAMD})$ & 0.771     & 0.912     & 0.975     & 0.982
              \\
    \noalign{\hrule height 0.5pt}
    \end{tabular}
    \end{center}
\end{table}

\begin{figure}[htbp]
    \centering
    \includegraphics[width=\figwidth]{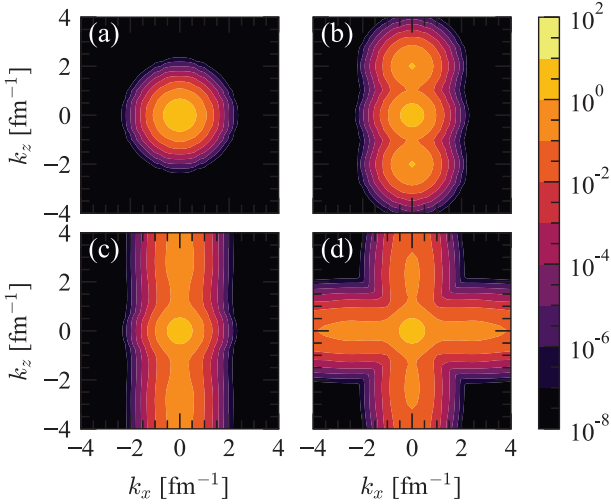}
    \caption{Momentum distributions of $^{4}$He nucleus described by the
      (HM-)AMD wave functions in the intrinsic frame. Panel (a) is the nucleon
      momentum distribution of the AMD basis without imaginary shifts
      ($\bm{D}=\bm{0}$). Panel (b) is the one for the HM-AMD basis with paired
      imaginary shifts ${D}_{z}=\pm5$ fm. Panel (c) is the nucleon momentum
      distribution of the superposed (HM-)AMD bases with various $z$-direction
      imaginary shifts $D_{z}$. Last Panel (d) is the nucleon momentum
      distribution of superposed (HM-)AMD bases with imaginary shifts in both
      $x$- and $z$-directions. Unit of the distribution is in fm$^{3}$.}
    \label{fig:intrinsic}
\end{figure}

Before showing the final results, we demonstrate the momentum distributions of
the $^{4}$He nucleus described by the HM-AMD wave functions in the intrinsic
frame. For the (HM-)AMD bases, $\bm{\textrm{Re}(Z)}=\bm{0}$ is commonly
adopted for all nucleons. For the AMD basis with $\bm{D}=\bm{0}$, the
spherical Gaussian distribution in the momentum space is obtained in
Fig.~\ref{fig:intrinsic}~(a). The effect of paired imaginary shifts
$D_{z}$=$\pm$5 fm in a HM-AMD basis of $^{4}$He is shown in
Fig.~\ref{fig:intrinsic}~(b), where additional two peaks of momentum
distribution is observed at corresponding $k_{z}=2\nu D_{z}=\pm$2.0 fm$^{-1}$.
Fig.~\ref{fig:intrinsic} ~(c) and (d) show the intrinsic nucleon momentum
distributions after superposing successively the (HM-)AMD bases with various
kinds of imaginary shifts in $z$- and $x$-directions, where the high-momentum
components beyond 2 fm$^{-1}$ are clearly shown as compared to the AMD basis
in Fig.~\ref{fig:intrinsic}~(a).

\begin{figure}[htbp]
    \centering
    \includegraphics[width=\figwidth]{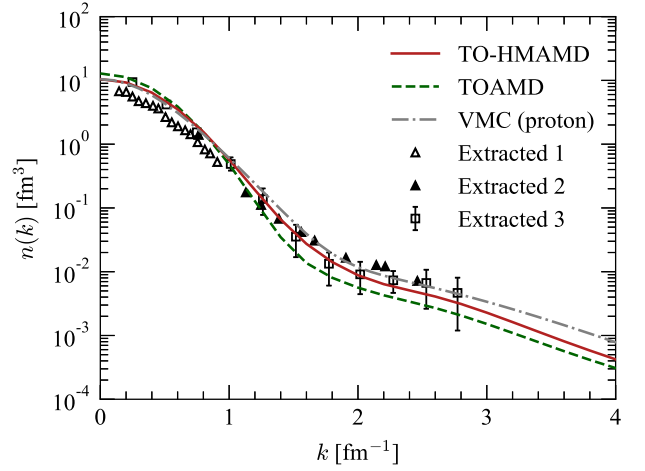}
    \caption{Nucleon momentum distribution of the $^{4}$He nucleus calculated
    using two-kinds of the wave functions of TO-HMAMD and TOAMD. The
    AV8$^\prime$ bare interaction is used to obtain the wave function. The
    dash-dotted curve and the extracted values are adopted from
    Ref.~\cite{ciofi96}. ``VMC (proton)'' denotes the proton
    distribution calculated with the Variational Monte Carlo wave function
    \cite{wiringa14}. The open squares represent the values extracted from
    inclusive $^{4}$He($e,e'$)$X$ reaction data \cite{ciofi91}. The full
    \cite{goff94} and open triangles \cite{brandt88} represent the values
    extracted from the exclusive $^{4}$He($e,e'p$)$X$ reaction data. The
    nucleon momentum distribution is normalized as $\int_{0}^{\infty}dk\,
    k^{2} n(k)=1$.}
    \label{fig:1f2f}
\end{figure}

We calculate the nucleon momentum distribution for the $^{4}$He nucleus using
both the TOAMD and the TO-HMAMD wave functions, and the results are compared
to the values extracted from experimental data in Fig.~\ref{fig:1f2f}. It is
found that both the TOAMD and the TO-HMAMD wave functions nicely describe the
high-momentum components, as shown by the solid curve and the dashed curve,
respectively. In particular, the TO-HMAMD wave function contains more
components of higher momentum and reproduces better the extracted values with
enhanced high-momentum tail, especially the inclusive data denoted by open
squares, as compared to the dashed curve.

\begin{figure}[htbp]
  \centering
  \includegraphics[width=\figwidth]{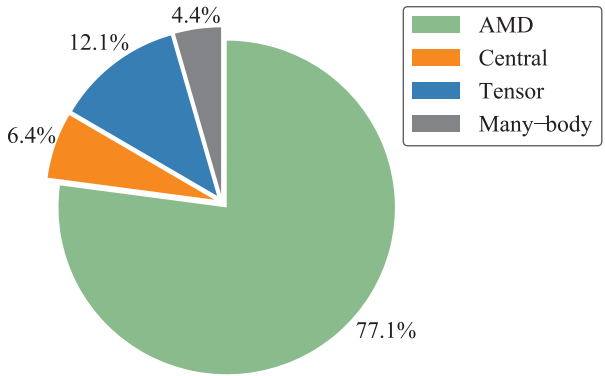}
  \caption{Probabilities of each orthogonal component of AMD, central, tensor,
  and many-body in the TO-HMAMD wave function of $^{4}$He.}
  \label{fig:pies}
\end{figure}

\begin{figure}[htbp]
    \centering
    \includegraphics[width=\figwidth]{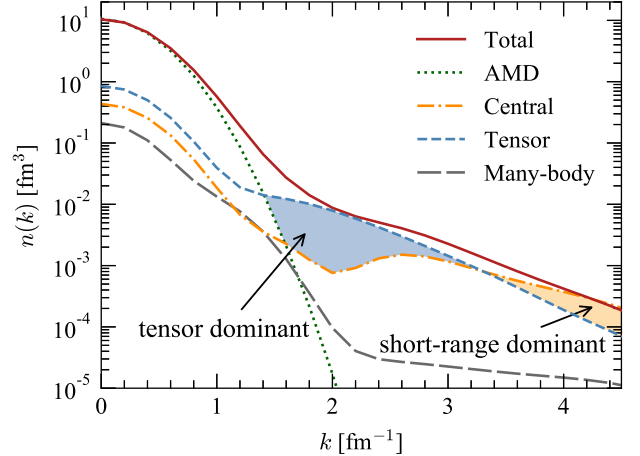}
    \caption{Decomposition of the high momentum component of the $^{4}$He
    nucleus. ``Total'' denotes the nucleon momentum distribution of the entire
    TO-HMAMD wave function. ``AMD'', ``Central'', and ``Tensor'' denote
    nucleon momentum distributions contributed by corresponding component
    defined in Eq.~(\ref{eq:components}).}
    \label{fig:components}
\end{figure}

We further explore the physical origins of the high-momentum components shown
in Fig.~\ref{fig:components} by the decomposition of the {\it ab initio} wave
function with respect to different channels of inter-nucleon correlations, and
demonstrate their contributions to the total momentum distribution. We prepare
first the following orthonormal wave functions $\ket{\Psi_{0\-- 3}}$ in the
coordinate space through the following Gram-Schmidt process
\begin{equation}\label{eq:components}
\begin{aligned}
  \textrm{\small{AMD:}}
&\ket{\Psi_{0}}=n_{0}\ket{\Psi_{\textrm{AMD}}},
  \quad\,
  \ket{\Psi_{S}}=n_{S}F_{S}\ket{\Psi_{\textrm{AMD}}},\\
  \textrm{\small{Central:}}
&\ket{\Psi_{1}}=n_{1}(1-
  \ket{\Psi_{0}}\bra{\Psi_{0}})
  \ket{\Psi_{S}},\\
  \textrm{\small{Tensor:}}
&\ket{\Psi_{2}}=n_{2}\,F_{D}\ket{\Psi_{0}},\\  
  \textrm{\small{Many-body:}}
&\ket{\Psi_{3}}=n_{3}(
  1
  -\ket{\Psi_{0}}\bra{\Psi_{0}}\\
&\phantom{\ket{\Psi_{3}}=n_{3}(1\,}
  -\ket{\Psi_{1}}\bra{\Psi_{1}}
  -\ket{\Psi_{2}}\bra{\Psi_{2}})
  \ket{\Psi}.\\  
\end{aligned}
\end{equation}
Here $\ket{\Psi_{0}}$ is the AMD wave function which corresponds to the
uncorrelated state. $\ket{\Psi_{1}}$ and $\ket{\Psi_{2}}$ are the two-body
$NN$-correlated terms in the central and tensor channels, respectively.
$F_{S}$ and $F_{D}$ are the two-body correlation operators adopted from the
TOAMD wave function in Eq.~(\ref{eq:toamd}) for the first order of cluster
expansion, and the central term $\ket{\Psi_{S}}$ contains the short-range
correlation at the two-body level. $\ket{\Psi_{3}}$ is the term of many-body
correlations which are not included in the TOAMD wave function. Coefficients
$n_{0,\dots,3,S}$ are normalization factors. Then, the {\it ab initio} wave
function $\ket{\Psi}$ can be expanded as a result of Eq.~(\ref{eq:components})
\begin{equation}\label{eq:wf-decomp}
    \ket{\Psi}=C_{0}\ket{\Psi_{0}}
    +C_{1}\ket{\Psi_{1}}
    +C_{2}\ket{\Psi_{2}}
    +C_{3}\ket{\Psi_{3}},
\end{equation}
where the coefficients $C_{i}=\braket{\Psi|\Psi_{i}}$ and the probabilities of
each term are given as $|C_{i}|^{2}$, as shown in Fig.~\ref{fig:pies}. In this
calculation, we found that the correlated terms contribute to about 23\% of
the total wave function in total, which is consistent with the estimation of
20\% in Ref.~\cite{hen17}. The contribution of tensor correlation term
$\ket{\Psi_{2}}$ is about 12\%, which is slightly smaller than the 13\% for
$D$-wave components in Ref.~\cite{myo17c}. This is reasonable because the
many-body term $\ket{\Psi_{3}}$ also contains small amount of $D$-wave
components. It is interesting to note that contribution from tensor
correlation is almost double as compared to the central contribution. The
many-body contribution of about 4\% is comparably small, although this term
contributes to about 10 MeV in the binding energy of $^{4}$He \cite{myo17a}.

Using the wave function in Eq.~(\ref{eq:wf-decomp}), we calculate the nucleon
momentum distribution for each component using the expansion method introduced
in Sec.~\ref{subsec:projection}, and the results are shown in
Fig.~\ref{fig:components}. In this figure, contributions from the AMD,
central, tensor, and many-body terms are clearly decomposed from the total
momentum distribution of the $^{4}$He nucleus, providing a different point of
view as compared to pioneer works in Refs.~\cite{alvioli12, alvioli13, neff15,
ryckebusch19}. It is found that the high-momentum components ($k>2$ fm$^{-1}$)
of $^{4}$He nucleus are mostly contributed by the tensor and the short-range
correlated terms, where the tensor correlation dominates around $k\approx 2$
fm$^{-1}$ and the short-range correlation dominates around $k\approx 4$
fm$^{-1}$, as denoted by the blue and yellow regions in the figure,
respectively. The contribution from many-body correlation term is found to be
comparably small for $k<4$ fm$^{-1}$ in this calculation. 

\begin{figure}[htbp]
  \centering
  \includegraphics[width=\figwidth]{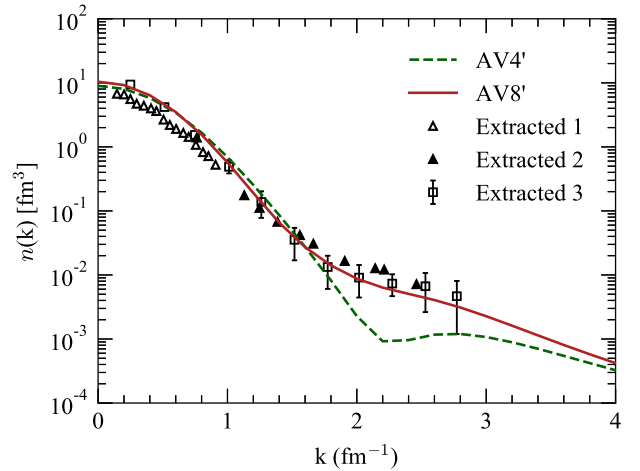}
  \caption{Nucleon momentum distribution of $^{4}$He calculated with the
  AV4$^\prime$ and the AV8$^\prime$ interactions in the TO-HMAMD framework.
  The values extracted from experimental data and normalization are the same
  as in Fig.~\ref{fig:1f2f}.}
  \label{fig:av48}
\end{figure}

Using the characteristics of tensor dominance around $k\approx 2$ fm$^{-1}$,
the tensor correlations in $^{4}$He can be confirmed by the shape of nucleon
momentum distribution, as shown in Fig.~\ref{fig:av48}. In this figure, we
compare two sets of momentum distributions predicted by using the AV4$^\prime$
and the AV8$^\prime$ interactions, with the values extracted from experimental
data. Both the AV4$^\prime$ and the AV8$^\prime$ interactions are renormalized
from the realistic AV18 interaction to reproduce the binding energies of light
nuclei \cite{pieper01b} and both interactions include strong short-range
repulsion in the central channel while AV4$^{\prime}$ does not have the tensor
and $LS$ force. In dashed curve with AV4$^{\prime}$, no tensor correlations
are included in the wave function. Hence, we observe a deep valley structure
around $k\approx 2$ fm$^{-1}$, which deviates significantly from the extracted
values. On the other hand, in the wave function using the AV8$^\prime$
interaction, the tensor correlations are induced by the $NN$ tensor force.
Consequently, a smooth solid curve is obtained and nicely reproduces the
extracted values. This comparison  provides a clear signature for the
validation of the tensor correlation in $^{4}$He nucleus.

\section{Conclusion}
We provided the fine-grained analysis for the different types of inter-nucleon
correlations in the $^{4}$He nucleus and investigated their contributions to
the high-momentum components. The nucleon momentum distribution of the
$^{4}$He nucleus was calculated by using the {\it ab initio} wave function in
the TO-HMAMD framework with the AV8$^{\prime}$ bare $NN$ interaction. The
first analytical formulation was derived for the nucleon momentum distribution
of (HM-)AMD wave functions, with complete subtraction of the center-of-mass
motion. In this work, the TO-HMAMD wave function was expanded by using the
HM-AMD bases using a new projection method, and it was found that a good
precision is obtained for the expansion. Based on these theoretical
preparations, we calculated the nucleon momentum distributions for the TOAMD
and the TO-HMAMD wave functions. It was found that both calculations predict
the high-momentum components and the distribution calculated from the TO-HMAMD
wave function reproduces the values extracted from experimental data with a
clear difference to the TOAMD case, which shows the effect of many-body
correlation. 

The physical origins of the high-momentum component were further clarified via
the decomposition of the total wave function into the orthogonal components
consisting of AMD, central, tensor and many-body channels, and their
contributions to the momentum distribution were calculated individually. The
tensor dominance around $k\approx$2 fm$^{-1}$ and the short-range dominance
around $k\approx$4 fm$^{-1}$ were observed from the decomposition. The
relatively small contribution from the many-body correlation was also
obtained. At last, the nucleon momentum distributions calculated by using both
the AV4$^\prime$ and the AV8$^\prime$ interactions were compared with the
extracted values, and the tensor correlation was found to be essential to
reproduce the smooth momentum distribution around $k\approx$2 fm$^{-1}$, which
provides a clear manifestation for the existence of tensor correlation in the
$^{4}$He nucleus. This work completes the previous understanding of
high-momentum component in the $^{4}$He nucleus, and is expected to be useful
for the theoretical and experimental studies in future for other heavier
nuclei.

\section*{Acknowledgments}
The authors would like to thank Prof.~Atsushi Hosaka and Prof.~Hiroki Takemoto
for the valuable discussions. M.L. acknowledges the support from the RCNP
theoretical group for his stay in RCNP and the fruitful discussions with the
members, and the support from the Yozo Nogami Research Encouragement Funding.
This work was supported by the JSPS KAKENHI Grants No. JP18K03660, and the
National Natural Science Foundation of China (Grants No. 11822503 and No.
11575082). The author N.W. is also supported by China Postdoctoral Science
Foundation (Grant No. 2019M661785). The present work is partially supported by
the Reimei Research Promotion project (Japan Atomic Energy Agency). The
numerical calculations were performed on the high performance computing server
at RCNP, Osaka University.

\def\JL#1#2#3#4{ {{\rm #1}} #2 (#4) #3}  
\nc{\PR}[3]     {\JL{Phys. Rev.}{#1}{#2}{#3}}
\nc{\PRC}[3]    {\JL{Phys. Rev.~C}{#1}{#2}{#3}}
\nc{\RMP}[3]    {\JL{Rev. Mod. Phys.}{#1}{#2}{#3}}
\nc{\PRA}[3]    {\JL{Phys. Rev.~A}{#1}{#2}{#3}}
\nc{\PRL}[3]    {\JL{Phys. Rev. Lett.}{#1}{#2}{#3}}
\nc{\NP}[3]     {\JL{Nucl. Phys.}{#1}{#2}{#3}}
\nc{\NPA}[3]    {\JL{Nucl. Phys.}{A#1}{#2}{#3}}
\nc{\PL}[3]     {\JL{Phys. Lett.}{#1}{#2}{#3}}
\nc{\PLB}[3]    {\JL{Phys. Lett.~B}{#1}{#2}{#3}}
\nc{\PTP}[3]    {\JL{Prog. Theor. Phys.}{#1}{#2}{#3}}
\nc{\PTPS}[3]   {\JL{Prog. Theor. Phys. Suppl.}{#1}{#2}{#3}}
\nc{\PTEP}[3]   {\JL{Prog. Theor. Exp. Phys.}{#1}{#2}{#3}}
\nc{\PRep}[3]   {\JL{Phys. Rep.}{#1}{#2}{#3}}
\nc{\PPNP}[3]   {\JL{Prog.\ Part.\ Nucl.\ Phys.}{#1}{#2}{#3}}
\nc{\JPG}[3]     {\JL{J. of Phys. G}{#1}{#2}{#3}}
\nc{\andvol}[3] {{\it ibid.}\JL{}{#1}{#2}{#3}}

\end{document}